%% file: main.tex
\documentclass[manuscript, screen, authorversion, acmsmall]{acmart}

\AtBeginDocument{%
  }

\setcopyright{acmlicensed}
\copyrightyear{2025}
\acmYear{2025}
\acmDOI{XXXXXXX.XXXXXXX}

\acmConference['26]{Make sure to enter the correct
  conference title}{xxxx xx--xx,
  2026}{}
\acmPrice{15.00}
\acmISBN{978-1-4503-XXXX-X/18/06}

\usepackage{multirow}
\usepackage{multicol}
\usepackage{pifont}%
\usepackage{bm}
\usepackage{xspace}

\definecolor{blue}{rgb}{0.2148,0.3412,0.8431}
\definecolor{yellow}{rgb}{0.7098,0.7216,0.0392}
\definecolor{green}{rgb}{0.10, 0.79, 0.68}

\newcommand\edit[1]{\textcolor{black}{#1}}

\newcommand\methodname{ChartOptimiser\xspace}
\begin{document}

\title{\methodname: Task-driven Optimisation of Chart Designs}

\author{Yao Wang}
\email{yao.wang@vis.uni-stuttgart.de}
\orcid{0000-0002-3633-8623}
\affiliation{%
  \institution{University of Stuttgart}
  \city{Stuttgart}
  \country{Germany}
}

\author{Danqing Shi}
\authornotemark[1]
\email{ds2206@cam.ac.uk}
\orcid{0000-0002-8105-0944}
\affiliation{%
  \institution{University of Cambridge}
  \city{Cambridge}
  \country{United Kingdom}
}
\affiliation{%
  \institution{Aalto University}
  \city{Espoo}
  \country{Finland}
}

\author{Jiarong Pan}
\authornote{Both authors contributed equally to this research}
\affiliation{%
  \institution{Bosch Center for Artificial Intelligence}
  \city{Stuttgart}
  \country{Germany}
}
\affiliation{%
  \institution{Eindhoven University of Technology}
  \city{Eindhoven}
  \country{Netherlands}
}

\email{garypan4@gmail.com}

\author{Zhiming Hu}
\email{zhiming.hu@vis.uni-stuttgart.de}
\orcid{0000-0002-5105-9753}
\affiliation{%
  \institution{University of Stuttgart}
  \city{Stuttgart}
  \country{Germany}
}

\author{Antti Oulasvirta}
\orcid{0000-0002-2498-7837}
\affiliation{%
  \institution{Aalto University}
  \city{Espoo}
  \country{Finland}
}
\affiliation{%
  \institution{ELLIS Institute Finland}
  \city{Espoo}
  \country{Finland}
}
\email{antti.oulasvirta@aalto.fi}

\author{Andreas Bulling}
\orcid{0000-0001-6317-7303}
\affiliation{%
  \institution{University of Stuttgart}
  \city{Stuttgart}
  \country{Germany}
}
\email{andreas.bulling@vis.uni-stuttgart.de}

\renewcommand{\shortauthors}{Wang et al.}

\begin{abstract}

\input{sections/abstract}

\end{abstract}



\begin{teaserfigure}
  \includegraphics[width=\textwidth]{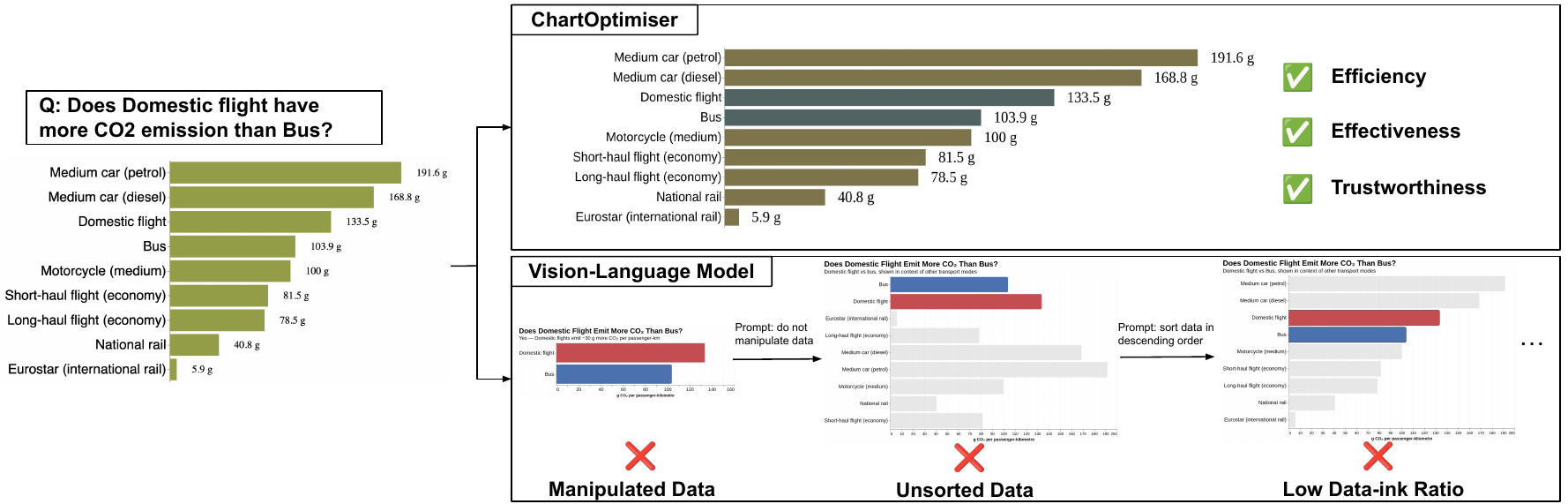}
  \caption{
  \edit{While Large-Language Models demonstrate broad competence, their application to visualisation often results in failures -- such as data manipulation, unsorted data, or low data-ink ratios -- unless guided by expert prompting. In contrast, \methodname guarantees efficient, effective, and trustworthy designs without requiring design expertise from users. By taking a plain chart and an analytical task as input (e.g. ``comparing CO$_2$ emissions''), \methodname employs Bayesian optimisation to effectively maximise a novel objective function composed of four perceptual metrics on visualisations.}
  }
  \Description{This work contributes ChartOptimiser that employs Bayesian optimisation to effectively maximise a novel objective function composed of four perceptual metrics.}
  \label{fig:teaser}
\end{teaserfigure}

\maketitle

\section{Introduction}

\input{sections/introduction}

\section{Related Work}\label{sec:relatedwork}

\input{sections/related_work}

\section{\NoCaseChange{\methodname}}\label{sec:method}
\input{sections/method}

\section{Dataset}\label{sec:dataset}
\input{sections/dataset}

\section{Evaluation}\label{sec:exp}
\input{sections/experiment}

\section{Discussion}
\input{sections/discussion}

\section{Conclusion}
\input{sections/conclusion}

\bibliographystyle{ACM-Reference-Format}
\bibliography{references}

\end{document}

%% file: sections/abstract.tex
\edit{
Automated chart design has seen significant advancements with the emergence of Large-Language Models (LLMs), which offer a practical solution for generating charts. 
However, LLMs frequently introduce possibly critical design failures, such as data manipulation and confabulation. 
While expert users can potentially mitigate these issues through iterative prompt engineering, this process requires substantial design knowledge and significant effort, remaining a massive barrier for the general public.
In this paper, we present \methodname, an automated method for generating chart designs with fidelity, efficiency, and effectiveness. 
Given the inter-dependencies between individual design parameters, \methodname employs Bayesian optimisation to effectively search the chart design space for a novel objective function grounded in four perceptual metrics. 
Our empirical evaluations in bar and pie charts demonstrate that \methodname eliminates iterative design loops, providing non-expert users with high-quality charts that outperform LLM-generated designs in chart clarity, task-solving ease, and visual aesthetics. 
}

%% file: sections/introduction.tex
The ability to effectively communicate information through visualisations is crucial in the information age~\cite{borner2019data, pozdniakov2023teachers}.
Well-designed charts not only satisfy viewers' information needs by enabling, e.g. easy retrieval or comparison of data values, but they can also enhance overall comprehension and decision-making processes~\cite{borkin2013makes, borkin2015beyond, yao2022visrecall}.
However, creating effective visualisations remains challenging and time-consuming -- visualisation designers must navigate a vast design space in which each design choice impacts others.

One challenge of chart design is the inter-dependencies between various design parameters, such as colour schemes, chart types, data scaling, or labels.
While the choice of parameters is crucial for the overall effectiveness of charts, these choices can also lead to misinterpretations if they are not made carefully.
The need to balance functional clarity with aesthetics further increases the difficulty of chart design.
Visual appeal can only be evaluated by ``taking a look at'' the chart; that is, viewers produce the image only when evaluating it.
Consequently, designers must understand visual perception principles and data characteristics to create understandable, informative, and effective visualisations.
Finally, it is challenging to design visualisations for the various analytical tasks that users may face when trying to satisfy their information needs~\cite{sedlmair2011information, ziemkiewicz2020open}.
Analytical tasks are among the most important factors influencing how viewers visually explore, make sense of, and extract information from visualisations.

The ever-increasing importance of data visualisations across various areas and the significant challenges in designing them call for automatic approaches that allow designers to optimise charts according to viewers' needs.
Some works focused on automatically \textit{generating} visualisations either using machine learning algorithms~\cite{chen2019towards, zhao2020chartseer} or rule-based constrains~\cite{wang_dracogpt_2024}.
These methods have been studied in various application contexts ranging from extracting reusable layout templates from visualisations~\cite{chen2020augmenting}, over generating visualisations from natural language~\cite{narechania2020nl4dv, luo2021natural}, to optimising the layout according to various metrics~\cite{micallef2017towards, wu2021learning}.
\edit{Recently, automated chart design has seen significant advancements with the emergence of Large-Language Models (LLMs), which offer a seamless solution for generating charts~\cite{tian2024chartgpt, wang_dracogpt_2024}. 
However, confabulation~\cite{siontis2024chatgpt}(also known as hallucination) is a longstanding problem that LLMs manipulates the data and compromises trustworthiness. 
Compared with the design constraints in rule-based chart optimisation methods, LLMs generate charts from billions of tokens that updated from ``in-the-wild'' data sources, leading to charts that may look plausible but violate chart design principles. 
While expert users can potentially mitigate these issues through iterative prompt engineering, this process requires substantial design knowledge and significant effort that remains a massive barrier for the general public.
} 

\edit{
To address these limitations, we propose a shift towards Bayesian optimisation to generate chart designs with the following advantages:
}

\begin{itemize}
    \item \edit{Fidelity: Bayesian optimisation searches within a constrained parameter space of valid chart specifications (e.g. VegaLite), guaranteeing data integrity and preventing data manipulation.}
    \item \edit{Efficiency: Bayesian optimisation reduces the manual effort of the iterative design revisions, such as prompt engineering to improve or correct LLM outputs.} 
    \item \edit{Effectiveness: While LLMs often focus on aesthetic plausibility, Bayesian optimisation directly maximises a novel objective function grounded in perceptual metrics, ensuring that the resulting visualisations are empirically effective for specific analytical tasks.}
\end{itemize}

Optimising visualisations to satisfy users' information needs is usually an iterative process that requires evaluating the visualisations at each iteration, which is inefficient and costly because the problem of finding the optimal parameter hypercube is inherently challenging, and the design space is vast.
Recently, Bayesian optimisation~\cite{shahriari2015taking} has shown good performance in optimising costly black-box problems and has been demonstrated to be highly effective for various applications such as optimising map searching~\cite{dudley2019crowdsourcing} and accelerating materials design~\cite{zhang2020bayesian}.
\edit{
However, it has not yet been explored for optimising chart designs to specific analytical tasks.
}

We propose \textit{\methodname} -- a novel Bayesian approach for task-driven optimisation of chart design that considers human visual attention.
At the core of \methodname is a novel objective function that combines four perceptual metrics to evaluate and automatically identify the optimal chart design parameters within an eight-dimensional design space: highlighting saliency, text legibility, colour preference, and white space ratio.
We demonstrate the usefulness of our method on the sample task of optimising the design parameters of bar charts -- a widely used chart type in data visualisation research and practice.
Through empirical evaluation of our method on 12 bar charts and four common visual analytical tasks -- find extremum, retrieve value, compare, and compute derived value -- we show that these four objectives are sufficient for effective optimisation, contrasting prior work that required many more objectives \cite{micallef2017towards}.
We further demonstrate that our \methodname outperforms existing design baselines concerning task-solving ease and clarity of the resulting chart. 
Our method also shows competitive performance with human-designed charts in terms of aesthetics and
surpasses baseline methods, particularly in more complex analytical tasks that involve comparison or computing derived values.
In summary, the contributions of our work are two-fold:
\begin{enumerate}
\item We introduce \methodname \,-- \edit{a fully} automatic method to optimise chart design parameters to users' analytical tasks. Our method uses Bayesian optimisation with a novel objective function that combines four perceptual metrics. 
We further showcased how \methodname could be adapted to many chart types such as multi-column bars and pie charts.

\item We report an empirical comparison of \methodname with four baseline design methods, including large-language models (e.g. ChatGPT), demonstrating our method's ability of fidelity, efficiency, and effectiveness in chart design optimisation.
\end{enumerate}

%% file: sections/related_work.tex
Our work is related to previous works on 1) Perceptual Metrics of Information Visualisation, 2) Automatic Visualisation Design, and 3) Visualisation Analytical Tasks.

\subsection{Perceptual Metrics of Information Visualisation}

It is indubitable that the effective use of colours is a key factor in any visual design, including information visualisations~\cite{rhyne2017applying, szafir2017modeling}. 
\citet{palmer2010ecological} proposed the weighted affective valence estimates (WAVE) score to quantify the colour harmony and aesthetics in visualisation design.
\cite{hegemann2023cocolor} proposed an integrated technique for colour exploration, assignment, and refinement, which took colour harmony, visual saliency, and elementary accessibility requirements into consideration.
Visual saliency, defined as a measure of the regions in a visual stimulus or scene that attract viewer attention, has been extensively studied, leading to the development of numerous effective saliency models in natural scenes~\cite{itti1998model, harel2007graph, kummerer2017deepgaze}. 
However, saliency models have been found to perform poorly when applied to information visualisations, as highlighted in prior work~\cite{matzen2017data, wang23_tvcg}. 
This triggered several research developing saliency models tailored for information visualisations in both free-viewing~\cite{matzen2017data, shin2022scanner, wang23_tvcg} and task-driven settings~\cite{wang2024salchartqa}. 
Some other perceptual metrics are specifically designed for information visualisations. 
Data-ink ratio quantifies the share of ink presenting data-related information in a visualisation~\cite{tufte1990envisioning}.
Visual density rates the overall density of visual elements without distinguishing between data and non-data elements~\cite{borkin2013makes}.
\citet{sedlmair2015data} evaluated how separable different classes were in projections of high-dimensional datasets.
Recallability and recognisability quantifies the quality of infographics design by testing how much information viewers remember after observation~\cite{borkin2015beyond, yao2022visrecall, wang2024visrecall++}.
\citet{micallef2017towards} proposed three perceptual metrics devised for scatterplots, including the perception of linear correlation, image quality, and classes and outliers.
As an integration of perceptual metrics, \citet{shin2023perceptual} proposed the Virtual Human Visual System, which provides accessibility, text, visual saliency, and visual density as feedback for iterative visualisation design, enabling designers to refine their visualisations better to align saliency with their design intentions.
Although the Virtual Human Visual System is promising, the involvement of visualisation designers is inevitable.
Instead, this paper builds a fully automatic design optimisation pipeline that incorporates four perceptual metrics into the objective function, \edit{without any human effort in the design loop}.

\subsection{Automatic Visualisation Design}
Automatic design systems can enumerate design candidates within the design space and identify the optimal layout to meet the needs of designers or users. 
It usually requires a utility or objective function to rank the design candidates.
Researchers have proposed many systems~\cite{moritz2018formalizing, hu2019vizml, luo2018deepeye, zhao2020chartseer} that recommend visualisations based on data structures and characteristics. 
Those systems focus on deciding the effective chart type, visual encoding, and data transformation. 
In addition to effectiveness, much research has optimised visualisations from other aspects. 
For example, \citet{hopkins2020visualint} addresses data integrity by surfacing chart construction errors such as truncated axes. 
Previous works have automatically extracted reusable layout templates from visualisations \cite{chen2020augmenting}, sketches~\cite{song2022vividgraph}, and infographics~\cite{lu2020exploring}. 
Other systems optimise the layout according to various metrics, such as similarities with user-input layouts~\cite{tang2020plotthread}, handcrafted energy terms, e.g. white space, scale, alignment, and balance~\cite{o2014learning}, perceptual metrics~\cite{micallef2017towards}, and crowdsourced layout quantifier~\cite{wu2021learning}. 
Generating visualisations from natural language (NL2Vis) has been a trend in automatic visualisation for many years~\cite{narechania2020nl4dv, luo2021natural, song2022rgvisnet,guo2024talk2data}.
NL2Vis takes a tabular dataset and a natural language query as input, specifying tasks and visualisation type, and generates visualisations that meet the requirement~\cite{narechania2020nl4dv, luo2021natural, cui2019text}. 
Recently, several works have demonstrated the superior performance of large language models (e.g. ChatGPT) for NL2Vis, highlighting their great potential in both in-domain and cross-domain settings~\cite{wu2024automated, chen2024viseval}.
However, current methods do not consider human perception feedback, such as visual attention.
This work optimises chart design for a combination of four perceptual metrics, making the optimised charts closer to observation conditions and human perception.

\subsection{Visualisation Analytical Tasks}
Strong evidence has shown that tasks significantly influence how viewers interact with visualisations~\cite{polatsek2018exploring, wang2024salchartqa}. 
\citet{amar2005low} identified 10 low-level analytical tasks (e.g., retrieving value, finding extremum, comparison) while another study~\cite{hibino1999task} highlighted high-level tasks such as background understanding, planning of analysis, and data exploration. 
\citet{saket2018task} evaluated the effectiveness of basic types of visualisations in the 10 low-level analytical tasks. 
\citet{elzer2011automated} proposed a Bayesian network to identify the communicative signals for intention recognition on simple bar charts. 
Eye movement data by \citet{polatsek2018exploring} and model by \citet{shi2025chartist} under three low-level analytical tasks demonstrated consistent gaze patterns within each task, with significant differences across tasks. 
Similarly, \citet{wang2024salchartqa} found saliency metrics are important indicators for visual analytical tasks.
Considering specific tasks is crucial, as visualisations can be designed to support a range of tasks based on the input data and can be evaluated by how effectively they enable task completion~\cite{schulz2013design}. 
To support task execution in visualisations, researchers have developed visualisations that explicitly highlight ``data facts'', such as trends or comparisons~\cite{srinivasan2018augmenting, shi2020calliope, shi2021autoclips, shi2024understanding}. 
In this work, we take task saliency as a metric in the objective function of the optimisation pipeline.

%% file: sections/method.tex
\begin{figure}[ht]
    \centering
    \includegraphics[width=\linewidth]{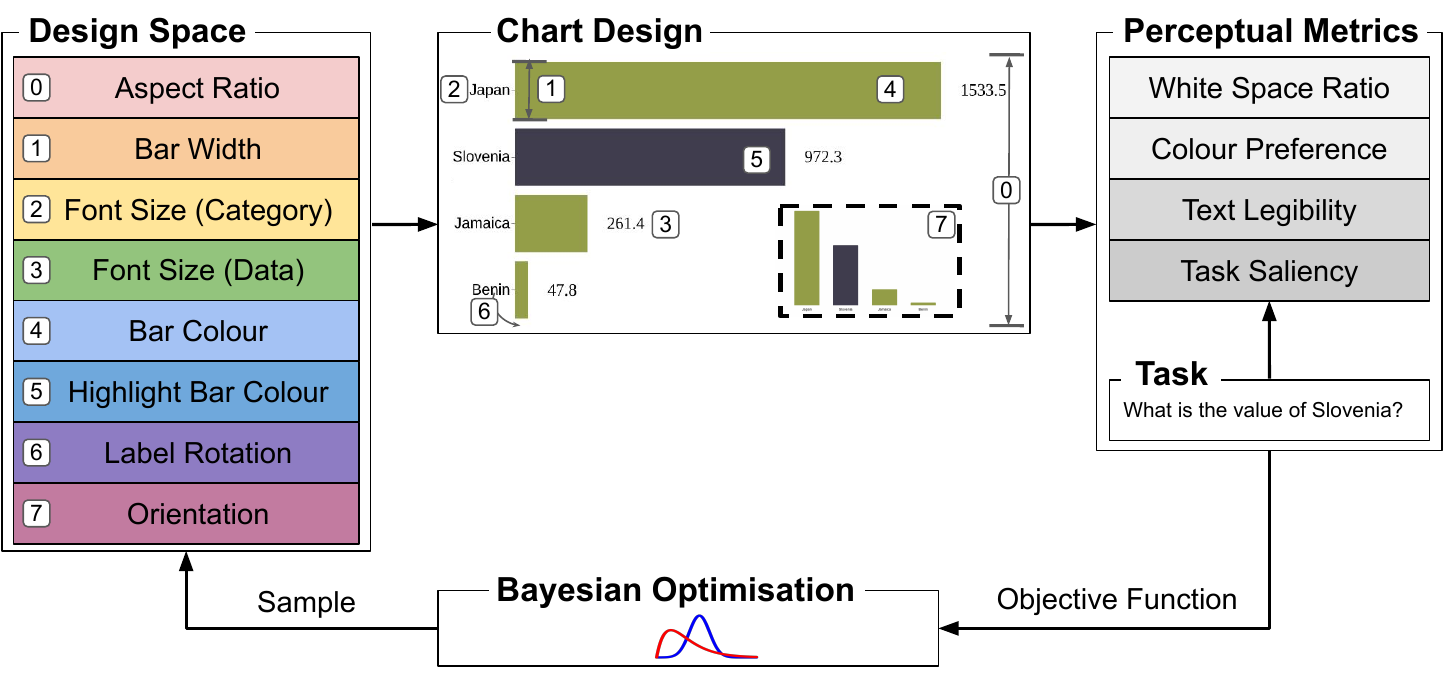}
    \caption{Overview of \methodname. Our method operates in a design space with eight dimensions, optimising it with respect to an objective function consisting of four perceptual metrics. Bayesian optimisation is used to sample from the design space and to maximise the objective function, thus optimising the chart design for a given visual analytical task.}
    \Description{Overview of \methodname. Our method operates in a design space with eight dimensions, optimising it with respect to an objective function consisting of four perceptual metrics. Bayesian optimisation is used to sample from the design space and to maximise the objective function, thus optimising the chart design for a given visual analytical task.}
    \label{fig:method}
\end{figure}

\methodname aims to automate the optimisation of chart design parameters for a given analytical task (see \autoref{fig:method}).
Our method operates on a design space consisting of eight dimensions and optimises these with respect to a novel objective function that combines four perceptual metrics.
Bayesian optimisation is used to sample from the design space and maximise the objective function.
In the following, we describe each of these components in more detail.
\edit{Our code is publicly available at \url{https://anonymous.4open.science/r/ChartOpt-96AB}}.

\subsection{Design Space}

The chart design space involves two main components~\cite{wu2021learning}: 1) \textit{visual encodings} that map data items to visual representations such as colour, position, and size, and 2) \textit{visual styles} that specify data-irrelevant visual designs like orientation and bar width. 
This work mainly focuses on bar charts, as they are the most widely used chart type for two-dimensional data. 
It divides the categorical data values into distinct bars, with values encoded by the bar length. 
Thus, categorical labels on axes associated with each bar are essential to a meaningful bar chart. 
Moreover, data labels on or next to bars showing the exact value of bars are also commonly used. 

For task-driven chart design, highlighting the bar\,(s) relevant to tasks was shown to facilitate task completion~\cite{elzer2011automated}.
For this initial work, we have limited ourselves to single-group bar charts to reduce the design space to a manageable scale.
Key parameters in the design space of bar charts are (see \autoref{fig:method}): 0) the aspect ratio of the chart, the font size of 1) axis labels and 2) data labels, 3) the bar width and 4) colour, often a 5) highlight bar colour, 6) the label rotation, and 7) the orientation of the chart. 
This set of design parameters extends the design space proposed in previous work~\cite{wu2021learning}.

We used the VegaLite~\cite{satyanarayan2016vega} grammar for interactive graphics to specify the data visualisations.
VegaLite specifications describe charts as encoding mappings from data to the visual properties of marks, which typically require a specific mark type and a set of one or more encoding definitions for the visual channels.
VegaLite expects a relational table of records with named fields. The mark type specifies the geometric objects used to encode data records visually. Possible values include bar, point, area, line, and tick. Visual encoding determines how data values map to the visual properties of marks. An encoding uses a visual channel such as spatial position, colour, size, shape, or text.
The VegaLite compiler will infer defaults based on the channel and data type if omitted. With the VegaLite descriptive grammar, the entire design space of charts is represented as

\begin{equation}
\mathcal{S} = V(\mathcal{X}_1 \times \mathcal{X}_2 ... \times \mathcal{X}_n),
\end{equation}

where $\mathcal{X}_1$ to $\mathcal{X}_n$ each denotes a dimension in the design space, such as aspect ratio and colour encoding. The $V$ function is the VegaLite compiler that automatically renders a chart from certain design parameters and determines the default properties for missing components based on a set of carefully designed rules~\cite{satyanarayan2016vega}. 
A certain parameter set $\mathbf{x} = (x_1, x_2,... x_n), x_i \in \mathcal{X}_i$ corresponds to one chart, denoted as $V_{\mathbf{x}}$.

\subsection{Objective Function}

The recent work~\cite{shin2023perceptual} demonstrated that including text legibility, visual saliency, and visual density as feedback can significantly facilitate the design process for designers. 
However, \citet{shin2023perceptual} ignored the influence of tasks on chart designs. 
As the effectiveness of visualisations varied among different analytical tasks\cite{saket2018task}, we aim to include a visual saliency metric biased by analytical tasks. 
As a conclusion, our novel objective function combines four perceptual metrics, providing task-specific objective in the automatic design loop.

The first metric $L_{w}$ is the \textit{White Space Ratio (WSR)}~\cite{o2014learning}, also known as area usage~\cite{wen2005optimization}. WSR is calculated as the ratio of pixels in $V_\mathbf{x}$ that are purely white (RGB value \#FFFFFF) to the total number of pixels in $V_\mathbf{x}$. 
This metric is critical for quantifying the visual density of chart designs~\cite{rosenholtz2007measuring}.
A high WSR suggests the presence of visual clutter, while a low WSR implies inefficient space utilisation.
The WSR distribution $\mathcal{N}(\mu, \sigma^2)$, derived from human-crafted charts in the ChartQA dataset~\cite{masry2022chartqa}, is considered the golden standard. Any WSR that deviates significantly from this standard is penalised \edit{for visually cluttered designs or leaving too much white space}.

\begin{equation}
  L_{w}(V_\mathbf{x}) =
    \begin{cases}
      0 & \text{ $WSR(V_\mathbf{x}) \in (\mu-\sigma, \mu+\sigma)$}\\
      - \left| WSR(V_\mathbf{x}) - \mu\right| & \text{otherwise}\\
    \end{cases}
\end{equation}

The second metric, \textit{Colour Preference ($L_{c}$)}, corresponds to the weighted affective valence estimates (WAVE) score \cite{palmer2010ecological}.
WAVE reflects human colour preference by calculating the average affective valence of people's responses to objects associated with specific colours, weighted according to the strength of the association with each colour.
Based on psychological studies, bright, saturated colours like red may evoke excitement or urgency, yielding a higher valence score. In contrast, softer colours like light blue yield a lower score.

\begin{equation}
    L_{c}(V_\mathbf{x}) = WAVE(V_\mathbf{x})
\end{equation}

The third metric is the \textit{Task Saliency ($L_{s}$)}, i.e. mean visual saliency in the task-related areas of interest (AOIs). 
This is inspired by previous works optimising graphical designs with visual importance maps~\cite{o2014learning, bylinskii2017learning, fosco2020predicting}. 
This term represents the average task-driven saliency value across the AOIs $task\_aoi_i(V_x)$. A well-designed chart that supports task-solving should effectively direct viewers' attention to these regions. Thus, a higher $L_s$ indicates more attention is drawn to the target region, suggesting better alignment with task objectives.

\begin{equation}
    L_{s}(V_\mathbf{x}) = \frac{1}{N} \sum^{i} Saliency(task\_aoi_i(V_\mathbf{x}))
\end{equation}

The fourth metric, \textit{Text Legibility ($L_{t}$)}, evaluates the readability of text in visual designs~\cite{shin2023perceptual}. A design fails if the text is not readable due to small font sizes, text overlap, or misalignment that extends beyond the image region. We employ Optical Character Recognition (OCR)~\cite{shi2016end} to automatically detect text characters in the rendered charts to approximate human perception. 
A penalty is applied when OCR fails to detect the text, indicating poor legibility.
For all $M$ categorical and data labels, the OCR score equals 1 if a label $text_m$ is successfully detected and equals 0 if missing. To ensure a smoother objective function, we construct a $P$-level image pyramid, where $downsample_p(V_\mathbf{x})$ is a downsampled version of the original image by a factor of $p$.

\begin{equation}
  L_{t}(V_\mathbf{x}) = \frac{1}{PM} \sum^{p} \sum^{m} OCR(downsample_p(V_\mathbf{x}), text_m)
\end{equation}

\begin{figure}[t]
    \centering
    \includegraphics[width=0.9\linewidth]{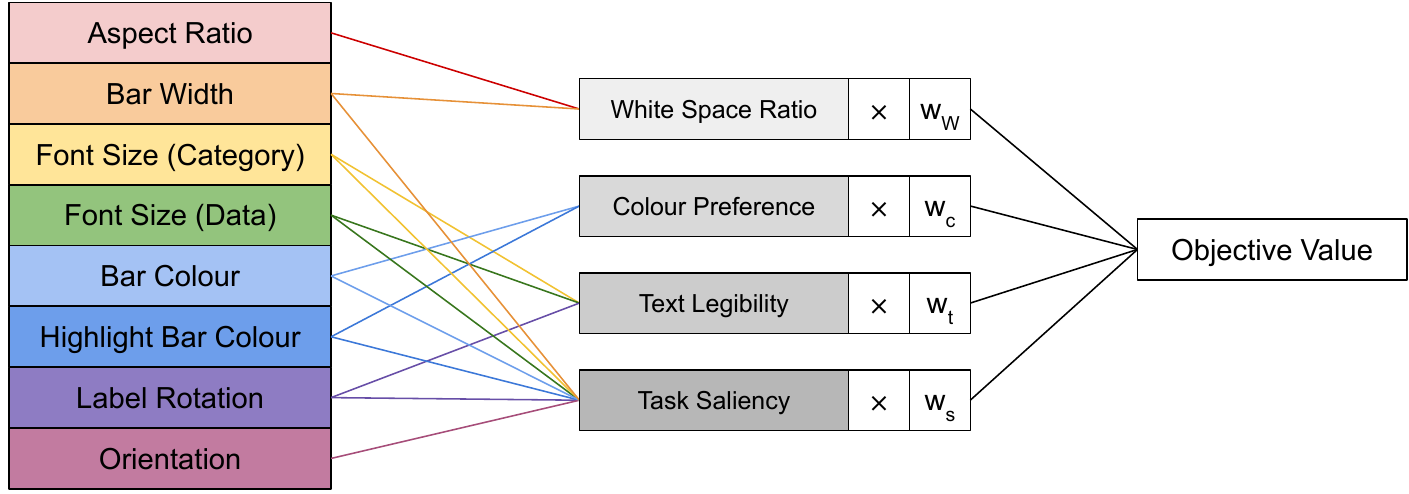}
    \caption{Associations between weighted perceptual metrics (centre) used in the objective function (right) and parameters in the design space of bar charts (left).}
    \Description{A visualisation of Associations between weighted perceptual metrics (centre) used in the objective function (right) and parameters in the design space of bar charts (left).}
    \label{fig:designparam_objfunc}
\end{figure}

All parameters in the design space are constrained by at least one metric in the objective function (see~\autoref{fig:designparam_objfunc}).
The \textit{WSR} controls the aspect ratio and bar width, preventing bar overlay or too far distances. 
The \textit{Colour Preference} metric controls bar colours, making the colours in the chart more harmonious.
The \textit{Task Saliency} metric generally influences chart layouts (orientation, aspect ratio, bar width) and colour, improving colour harmony and bar distance.
Finally, the \textit{Text Legibility} constrains the font size of the category and data labels.
For example, optimising for \textit{Text Legibility} may influence the font size of categorical and data labels and the rotation of these labels, while optimising for \textit{WSR} may influence the width and aspect ratio of the bars themselves. Optimising for both may influence all of these metrics jointly. 

The final objective function $L$ is 

\begin{equation}
    L(V_\mathbf{x}) = w_{w} L_{w}(V_\mathbf{x}) + w_{c} L_{c}(V_\mathbf{x}) + w_{t} L_{t}(V_\mathbf{x}) + w_{s} L_{s}(V_\mathbf{x})
\end{equation}

\subsection{Bayesian Optimisation}

Evaluating charts poses significant challenges as identifying the parameter hypercube that yields the optimal value of the objective function is time-consuming. 
The challenges are threefold. 
First, identifying a ``golden standard'' for a specific design parameter is challenging since the interdependencies between parameters are inherently complex.
Second, despite using descriptive grammars like VegaLite, the design space $\mathcal{S}$ remains large, encompassing continuous chart design parameters such as bar width and colour. 
Third, each evaluation requires generating a visual representation of the chart, as all terms in the objective function depend on rendered images. 
For instance, saliency maps could only be predicted once images were rendered. 
Repeated image generation across iterations incurs substantial computational costs. These factors underscore the need for sample-efficient optimisation techniques to reduce the number of costly evaluations required to achieve an optimal layout.

Inspired by \cite{dudley2019crowdsourcing} that used Bayesian optimisation to refine 2D interface design in a map searching use case, we built a BO pipeline to balance efficiency and effectiveness. 
Bayesian optimisation \citep{shahriari2015taking} is a popular approach to optimise costly functions across various applications, which typically employs a probabilistic surrogate model to approximate the black-box function and estimate uncertainty. 
An acquisition function then utilises these estimates to determine the next query point. 
\edit{The non-parametric and sample-efficient nature of Gaussian Process fits our design optimisation problem well.}
Therefore, we use a Gaussian process (GP) \citep{williams2006gaussian} as the surrogate model, which uses a kernel or covariance function $k$ to encode the prior belief for the smoothness of the black-box function $f$.

Let the perceptual metrics $f = L \circ Vega$ be the black-box function to optimise, we perform a sequential optimisation to find the optimal design parameters $\mathbf{x}^{*} \in \mathrm{arg\,max}_{\mathbf{x} \in \mathcal{X}_1 \times \mathcal{X}_2 ... \times \mathcal{X}_n}$\, $ f(\mathbf{x}) = \mathrm{arg\,max}_{V_{\mathbf{x}} \in \mathcal{S}}\, L(V_{\mathbf{x}})$ based on the previously collected data, where in each iteration, we collect a new noisy observation $y_i \sim \mathcal{N}(f(\mathbf{x}_i), \sigma^2)$. Given data $\mathcal{D}_n = \{(\mathbf{x}_{i}, y_{i})\}_{i=1}^{n}$ that are collected up to $n$-th iteration during optimisation, the posterior prediction of a GP $p(f(\mathbf{x}) \mid \mathcal{D}_n)$ at a test location $\mathbf{x}$ is specified by the mean function $\mu_n(\mathbf{x})$ and the covariance function $v_n(\mathbf{x})$:
\begin{equation}
    \mu_n(\mathbf{x}) = \mathbf{k}_n(\mathbf{x})^\mathsf{T}(\mathbf{K}_n + \sigma^2 \mathbf{I})^{-1}\mathbf{y}_n \,, \qquad
    v_n(\mathbf{x}) = k(\mathbf{x}, \mathbf{x}) - \mathbf{k}_n(\mathbf{x})^{\mathsf{T}}(\mathbf{K}_n+\sigma^2 \mathbf{I})^{-1}\mathbf{k}_n(\mathbf{x})\,,
\end{equation}
where $[K_n]_{ij} = k(\mathbf{x}_i, \mathbf{x}_j) $, $\mathbf{k}_n = [k(\mathbf{x}, \mathbf{x}_1), \cdots, k(\mathbf{x}, \mathbf{x}_n)]^{\mathsf{T}}$ %
and $\sigma$ is the variance of the noise.  $k(\mathbf{x}, \mathbf{x'})$ is the similarity prediction based on a type of kernel $k$. We use radial basis function kernel in the experiment.
$\mathbf{k}_n$ is the vector containing all the similarity measures between input point x and all other existing data $x_1, ..., x_n$.

With the posterior prediction, we can compute the utility of a given point $\mathbf{x}$ via an acquisition function $\alpha(\mathbf{x})$. One of the most popular acquisition functions is the Expected Improvement (EI) \citep{mockus1974bayesian}:
\begin{equation}
    \alpha^{\mathrm{EI}}(\mathbf{x};\mathcal{D}_n) = \mathbb{E}_{f(\mathbf{x})} [\max(f(\mathbf{x}) - y^{*}, 0)]= \int \max(f(\mathbf{x}) - y^{*}, 0) p(f(\mathbf{x}) \mid \mathcal{D}_n) \mathrm{d}f(\mathbf{x})\,,
\end{equation}
where $y^{*}$ is denoted as the best-observed value till step $n$. We can make a recommendation for the next iteration by maximising the acquisition function 
\begin{equation}
x_{n+1} \in \mathrm{arg\,max_{\mathbf{x} \in \mathcal{X}_1 \times \mathcal{X}_2 ... \times \mathcal{X}_n}\, \alpha^{\mathrm{EI}}(\mathbf{x};\mathcal{D}_n}).
\end{equation}

In summary, we use BO to find the optimal design parameters by treating the user's perceptual behaviour as a black-box function. 
We use a Gaussian Process~\citep{williams2006gaussian} to model the joint perceptual objective $L$ based on four perceptual terms described in section 3.2. 
The chart design parameters are optimised by maximising the acquisition function iteratively.

\subsection{Implementation Details}

\paragraph{Specification of Objective Functions}
\edit{For white space ratio, the normal distribution has $\mu = 0.496, \sigma = 0.263$.} 
The implementation of the WAVE algorithm for colour preference was based on the Aalto Interface Metrics platform~\cite{oulasvirta2018aalto}\footnote{\url{https://github.com/aalto-ui/aim}}. 
For task saliency, task-relevant regions were automatically detected from the rendered chart in SVG format. 
For the FE and RV tasks, one task-related region was identified. 
In contrast, two or more target regions were defined for the CDV and CP tasks. 
To align these regions with annotated questions or answers, we matched the ``aria-label'' attribute in the SVG with the corresponding elements. 
We utilised the \textit{boundingbox()} function from \textit{svg.path}\footnote{https://pypi.org/project/svg.path/} to extract the task-related bounding boxes.
The saliency maps were generated using the VisSalFormer model~\cite{wang2024salchartqa}. 
Regions with zero saliency were excluded from the analysis by discarding pixels within bounding boxes that had a saliency value of 0. 
For text legibility assessments, we applied Optical Character Recognition (OCR), using PyTesseract\footnote{\url{https://pypi.org/project/pytesseract}} on a three-level image pyramid (at 1/8, 1/4, and 1/2 of the original image size).

\paragraph{Parameter Boundaries}
The bounds for each parameter were carefully selected to define the limits of the hypercube. 
We set the aspect ratio between 0.33 and 3, varying the chart width from 200 to 1,800 pixels. 
Font sizes were restricted to 10-36 pixels, bar width to 20-180 pixels, and label rotation to \{0, -45$^\circ$, -90$^\circ$\}. 
We permitted the full range of the HSV space for bar colours.

\paragraph{Training Details}
\edit{The BO pipeline is built with the BoTorch\footnote{\url{https://botorch.org/}} and Adaptive Experimentation Platform\footnote{\url{https://ax.dev/}}.}
Weights of the objective function were set to $w_w = 3$, $w_c = 1$, $w_t = 2$, $w_s = 4$.
The candidate list was constructed by sampling from the parameter hypercube using a Sobol sequence, \edit{with 16 trials generated and a minimum of 5 trials completed in each step}.
We sampled 50 candidates with \methodname for every chart-task pair, which took about 3-6 seconds per step. 
While more candidates provide better search resolution, we considered 50 candidates sufficient to demonstrate the strength of \methodname.
All experiments were conducted on a Desktop PC with an AMD Ryzen 7 3700K 8-Core Processor and NVIDIA GeForce RTX 2060 Super GPU with 8GB VRAM.

%% file: sections/dataset.tex
\paragraph{Data Preparation}
Implementing the proposed pipeline requires a vector-based chart dataset that supports customisable chart design parameters and includes task-based optimisation goals. 
We utilised the ChartQA dataset~\cite{masry2022chartqa}, a chart question-answering dataset comprising real-world charts and high-quality human-annotated questions. 
Each chart in ChartQA is associated with two human-annotated analytical tasks. We restricted our selection to single- \edit{and multi-column} bar charts to simplify the design space. 
Since bar colours were changed in the optimisation pipeline, we filtered out visual questions about bar colours, such as ``What is the value of the yellow bar?''.
The final selection comprised 300 single-column bar charts (207 horizontal and 93 vertical) and \edit{258 multi-column bar charts (all horizontal)}. These charts were recreated using the VegaLite grammar~\cite{satyanarayan2016vega}\footnote{\url{https://vega.github.io/vega-lite}}, which allowed for declarative control over chart properties via key-value pairs in JSON format.
This reduces the design space to a reasonable scale, consistent with previous automatic design literature~\cite{moritz2018formalizing, wu2021learning}.
We created templates \edit{for single-column and multi-column bar charts} with fixed design parameters in VegaLite and filled the JSON template with data values from the ChartQA annotations. 
In line with previous work~\cite{wu2021learning}, the chart height was fixed at 600 pixels, while the width was initialised to 600 pixels.
Bar widths were all initiated to 40 pixels, and \edit{data labels and axis labels were uniformly set to a font size of 24 and 17 pixels, respectively.} 
\edit{For single-column bars, all bars were assigned the colour \#949d48, a common hue in ChartQA. 
For multi-column bars, we used the default colour palette in VegaLite.}
The orientation of each chart followed the original designs in ChartQA. 
The VegaLite compiler then rendered the JSON files into Portable Network Graphics (PNG) and Scalable Vector Graphics (SVG) formats, \edit{which were generated by Vega-Altair~\footnote{https://altair-viz.github.io/}}.

\paragraph{Tasks}
Performing analytical tasks is a common use case when people read charts~\cite{elzer2011automated, polatsek2018exploring}. Effective chart designs should enable viewers to easily perform analytical tasks, facilitating efficient communication of information. 
In this work, four common analytical tasks were studied (see \autoref{fig:teaser}):
 
\begin{itemize}
    \item \textit{Find Extremum (FE)}~\cite{amar2005low}. FE tasks require labelling a data point with an extreme value of an attribute.
    Example task: Which mode of transport has the fewest CO2 emissions? 
    To solve this task, participants must find the shortest bar in the chart and read the corresponding categorical label \textit{Eurostar (international rail)}.

    \item \textit{Retrieve Value (RV)}~\cite{amar2005low}. RV tasks ask for the data value given a specific target.
    Example task: \textit{What is the value of Medium car (petrol)?} 
    To solve this task, participants should look up the categorical labels to find Medium car (petrol), then read the data label for the value \textit{191.6\,g}.

    \item \textit{Compute Derived Value (CDV)}~\cite{amar2005low}. CDV tasks require participants to perform RV tasks to retrieve the value of given data points and then calculate the derived value. 
    Example task: \textit{What is the average value of National rail and Bus?} 
    To solve this task, participants should look up the National rail and Bus values and then calculate the average of these two values.

    \item \textit{Compare (CP)}~\cite{albers2014task}. CP tasks require participants to perform RV tasks to retrieve the value of given data points (usually two) and then make judgements about data properties. 
    Example task: \textit{Does Domestic flight have more CO2 emission than Bus?} 
    To solve this task, participants should look up the values for domestic flights and buses and then determine whether domestic flights have higher CO2 emissions.
\end{itemize}

%% file: sections/experiment.tex
To gain further insights, we conducted a user study to let participants qualitatively rate chart designs generated from our method against human-designed charts and three strong baseline approaches.
Then, we further analysed the optimised parameters from \methodname, and showcased how \methodname could be applied to other chart types, including multi-column bar and circular charts.

\subsection{Study Design}
\label{ssec:studydesign}
\paragraph{Participants}

We recruited 60 participants (31 males, 29 females) aged from 20 years (M\,=\,33.75, SD\,=\,10.94) via the Prolific platform. 
Only participants with normal or corrected-to-normal vision were involved. 
57 of the participants confirmed they understood how to read bar charts, and 45 had experience designing bar charts before. 
The study took an average of 34 minutes (SD\,=\,19\,minutes) to complete.
Participants received 5 pounds as compensation. 
Upon completing all trials, we asked participants to provide qualitative feedback on the most important characteristics they used in their subjective evaluation.

\paragraph{Experimental Protocol}

\begin{figure}[t]
    \centering
    \includegraphics[width=\linewidth]{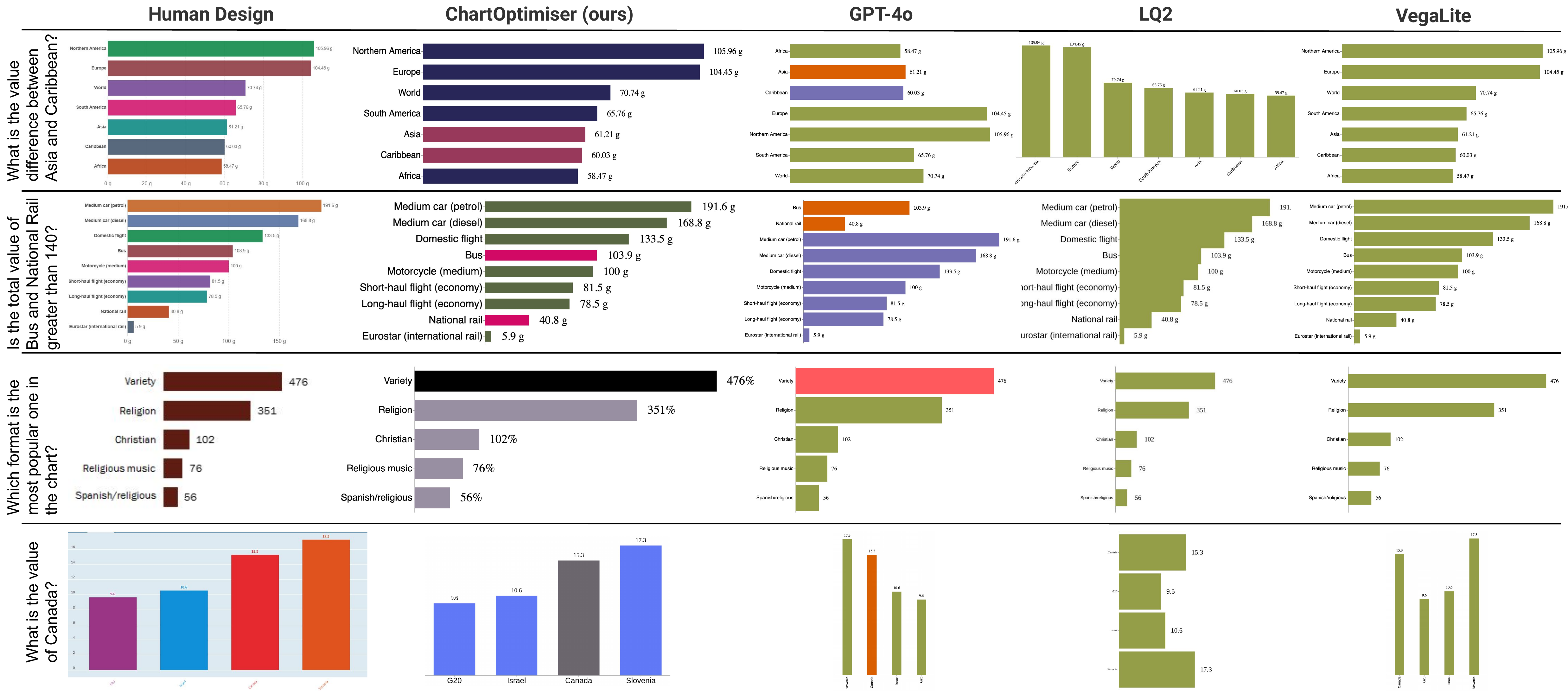}
    \caption{From left to right: Sample bar charts from the ChartQA \cite{masry2022chartqa} dataset and corresponding bar charts optimised using \methodname, GPT-4o, LQ2~\cite{wu2021learning}, as well as VegaLite default.}
    \Description{Sample bar charts from the ChartQA dataset, optimised bar charts from \methodname, LQ2, and GPT-4o, as well as VegaLite default.}
    \label{fig:chart_baselines}
\end{figure}

We randomly sampled 12 charts from the ChartQA dataset, corresponding to three tasks from each of the four task types introduced in Section 3: Find Extremum (FE), Retrieve Value (RV), Compute Derived Value (CDV), and Compare (CP).
RV and FE were considered simple tasks, as they require observing only a single data point. 
In contrast, CP and CDV were considered more complex, necessitating observing multiple data points and performing calculations~\cite{elzer2011automated}.
In the study, participants were introduced to the tasks before starting.
We presented participants with five chart designs in each of the 12 study tasks (see \autoref{fig:chart_baselines}).
Study participants were asked to evaluate the five chart designs from three aspects: 
1) chart aesthetics,
2) chart clarity,
3) and task-solving ease.
Each aspect was evaluated on a 7-point Likert scale, where 1 is the strongest negative, and 7 is the strongest positive.
After completing ratings for each task, participants are asked to provide brief comments explaining their ratings.

\paragraph{Design Comparison}

We compared the chart designs from \methodname against human-designed charts and three baseline approaches. Several examples of chart designs are shown in \autoref{fig:chart_baselines}.

\begin{itemize}
    \item \textit{Human-designed charts (Human)}. These are human-designed bar charts from the original ChartQA dataset~\cite{masry2022chartqa}. We cropped the additional information in the image to preserve only the data regions for a fairer comparison.
    \item \textit{VegaLite default designs (VegaLite)}. These are the recreated charts from ChartQA data tables, using the VegaLite library~\cite{satyanarayan2016vega} with fixed chart design parameters (see Section 4.2).
    \item \textit{GPT-4o optimised charts (GPT-4o)}. These are the responses from \edit{the API of} GPT-4o\footnote{\url{https://platform.openai.com/docs/models/gpt-4o}}~\cite{achiam2023gpt} with the prompts ``\edit{\textit{You are an expert in data visualisation. Given the following VegaLite chart specification, suggest improvements to enhance chart clarity and visual aesthetics. 
    VegaLite Specification:}} \{\$VEGA\_JSON\}'', and ``\textit{also optimise the above Vegalite JSON with the task:} \{\$TASK\}, \edit{\textit{respond only with the new VegaLite Specification in JSON format}}''. The \{\$VEGA\_JSON\} is the JSON file of the VegaLite baseline, while the \{\$TASK\} is either FE, RV, CP, or CDV task from the ChartQA dataset. We noticed that GPT-4o filters data in some CP and CDV tasks, so we added the prompt ``\textit{do not change the data, add the filtered data back}'' to prevent editing the data source. 
    \item \textit{LQ2 optimised charts (LQ2)}. The Layout Quality Quantifier (LQ2)~\cite{wu2021learning} is a machine learning model that evaluates chart layouts from paired crowdsourcing data, yielding a preference score given the layout parameters as input. We used the LQ2 score as the objective function in our proposed Bayesian pipeline to optimise charts. It takes the number of bars, bar width, and aspect ratio (with height fixed) as input and yields a score ranging from 0 (worst) to 1 (best). Since the LQ2 score does not constrain colour, we fixed the design parameter of dimension 4) the bar colour and 5) the highlight bar colour for LQ2. The VegaLite baseline was used as the initial parameter of LQ2.
\end{itemize}

\subsection{Study Results}

\begin{figure}[t]
    \centering
    \includegraphics[width=0.9\linewidth]{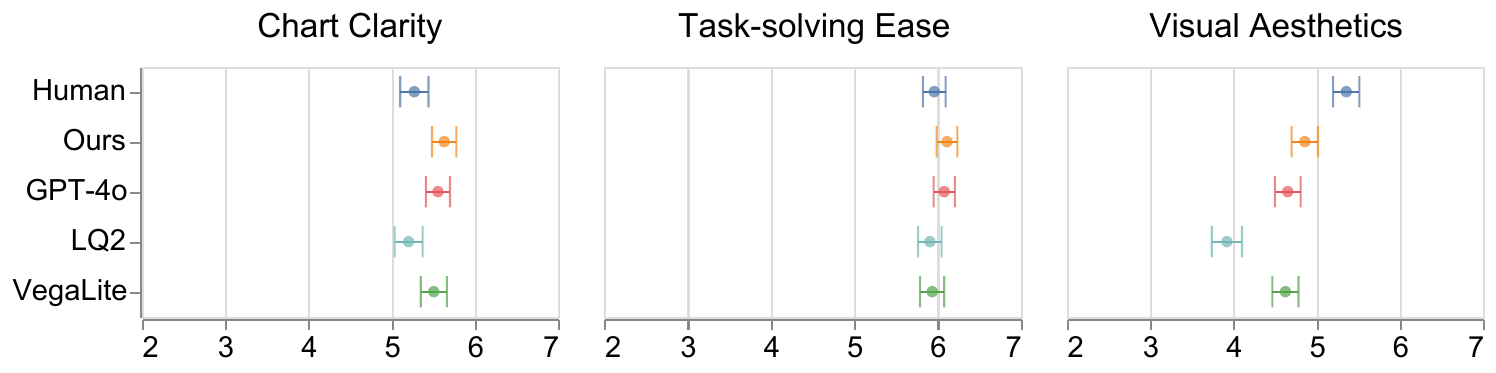}
    \caption{Mean ratings and 95\% confidence intervals for visual aesthetics, chart clarity, and task-solving ease for the different methods.}
    \Description{Mean ratings and 95\% confidence intervals for aesthetics, clarity, and task-solving ease for different methods.}
    \label{fig:overall-rating}
\end{figure}

\paragraph{Ratings}
\autoref{fig:overall-rating} illustrates the participants' ratings for visual aesthetics, chart clarity, and task-solving ease for all chart designs. Overall, \methodname ranks first in clarity ($M=5.6, SD=1.4$) and task-solving ease ($M=6.1, SD=1.2$), with the GPT-4 design coming in second. The human-designed chart outperforms others in aesthetics ($M=5.4, SD=1.5$), with our \methodname ranking second ($M=4.9, SD=1.5$). 
The results indicate that human designers consider more details to ensure the charts are pleasing, while automatic approaches can improve chart clarity and task-solving ease.
The normality of the ratings was confirmed via the Shapiro-Wilk Normality test.
A one-way ANOVA revealed a statistically significant difference in ratings between these methods in aesthetic ($F(4, 1795)=39.2, p < 0.01$) and clarity ($F(4, 1795)=3.2, p < 0.01$).
The task-solving ease was insignificant based on the one-way ANOVA analysis ($F(4, 1795)=1.8, p = 0.13$).
No significant differences were found in the post-hoc Tukey’s HSD test between methods regarding task-solving ease ($p > 0.05$). However, in terms of aesthetics, human designs were significantly better than other methods ($p < 0.05$). Additionally, \methodname was significantly better than LQ2 ($p < 0.05$) and slightly better than GPT-4o and VegaLite ($p > 0.05$). In clarity, \methodname was significantly better than human designs and LQ2 ($p < 0.05$), but also slightly higher than GPT-4o and VegaLite ($p > 0.05$). 
This result suggests that the method has the potential to design charts that are both aesthetically pleasing and clear for tasks.

\begin{figure}[t]
    \centering
    \includegraphics[width=0.9\linewidth]{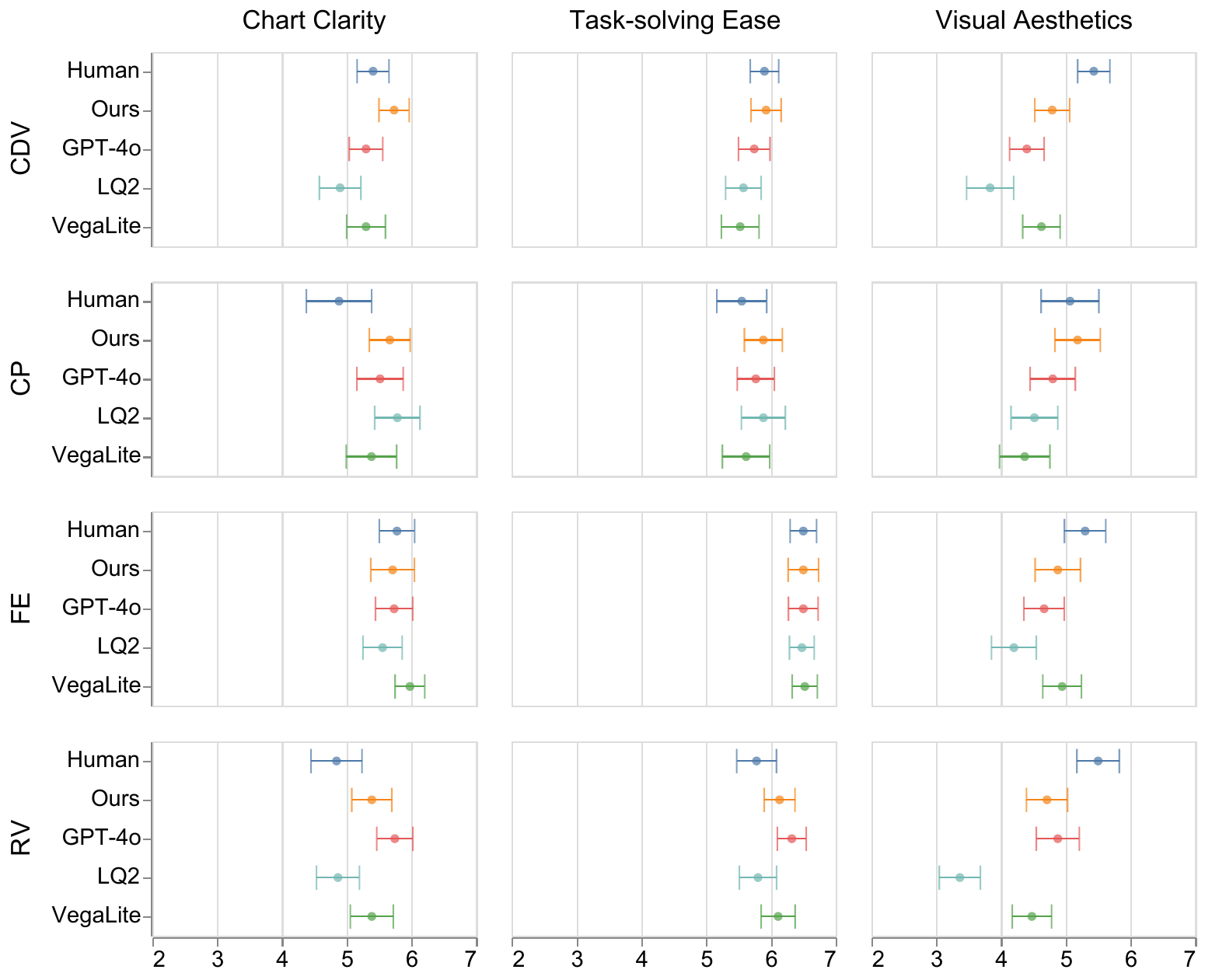}
    \caption{Mean ratings and 95\% confidence intervals for visual aesthetics, chart clarity, and task-solving ease for the different methods and four task types.}
    \Description{Mean ratings and 95\% confidence intervals for aesthetics, clarity, and task-solving ease on different methods over four task types.}
    \label{fig:task-rating}
\end{figure}

Next, we examined the ratings for each task (see \autoref{fig:task-rating}):
1) \textit{CDV}: \methodname outperformed others in clarity and task-solving ease, ranked second in aesthetics, following human-designed charts.
2) \textit{CP}: \methodname ranked first in aesthetics, surpassing human designs that ranked second. This is the only task for which human designs did not rank first. Our method ranked in the top two for clarity and task-solving, with no significance found against LQ2, which ranked first ($p > 0.05$).
3) \textit{FE}: All designs received high task-solving ratings, suggesting this task's easiness. VegaLite ranked first in clarity, as its simplest design can clearly display extreme values. Human design ranked first in aesthetics.
4) \textit{RV}: GPT-4o designs performed well in the simple retrieve value task, with \methodname in second place with no significance.
Our proposed approach performed best in complex tasks (CDV and CP). It also demonstrated its superiority in simple tasks, although comparable to other approaches.

\paragraph{Comments}

Participants' comments indicated that our objective function's design was effective by emphasising saliency, text legibility, colour preference, and white space ratio.
Participants frequently mentioned the following factors that influenced their ratings:
1) \textit{Highlighting key data points (mentioned 17 times).} Using different colours to emphasise the highest value or relevant data points made it easier for users to quickly identify important information;
2) \textit{Clear labelling (mentioned nine times).} Easily visible labels and numbers next to each bar were essential for quickly understanding the data and making decisions based on it;
3) \textit{Pleasing colour scheme (mentioned five times).} Aesthetically pleasing colour combinations made charts more engaging, especially using contrast or harmonious colours. Charts with multiple colours were generally preferred over monochromatic designs;
4) \textit{Well-spaced bars (mentioned four times).} Well-spaced bars and well-proportioned bar sizes enhanced visual appeal and improved the overall aesthetic experience. Users preferred charts that had clear distinctions between different data segments;
5) \textit{Legible font sizes (mentioned three times).} Readable font sizes for labels and values were essential for making charts easily understandable. Smaller fonts were noted as a barrier to clarity.

\subsection{Parameter Analysis}

\begin{figure}[t]
    \centering
    \includegraphics[width=0.9\linewidth]{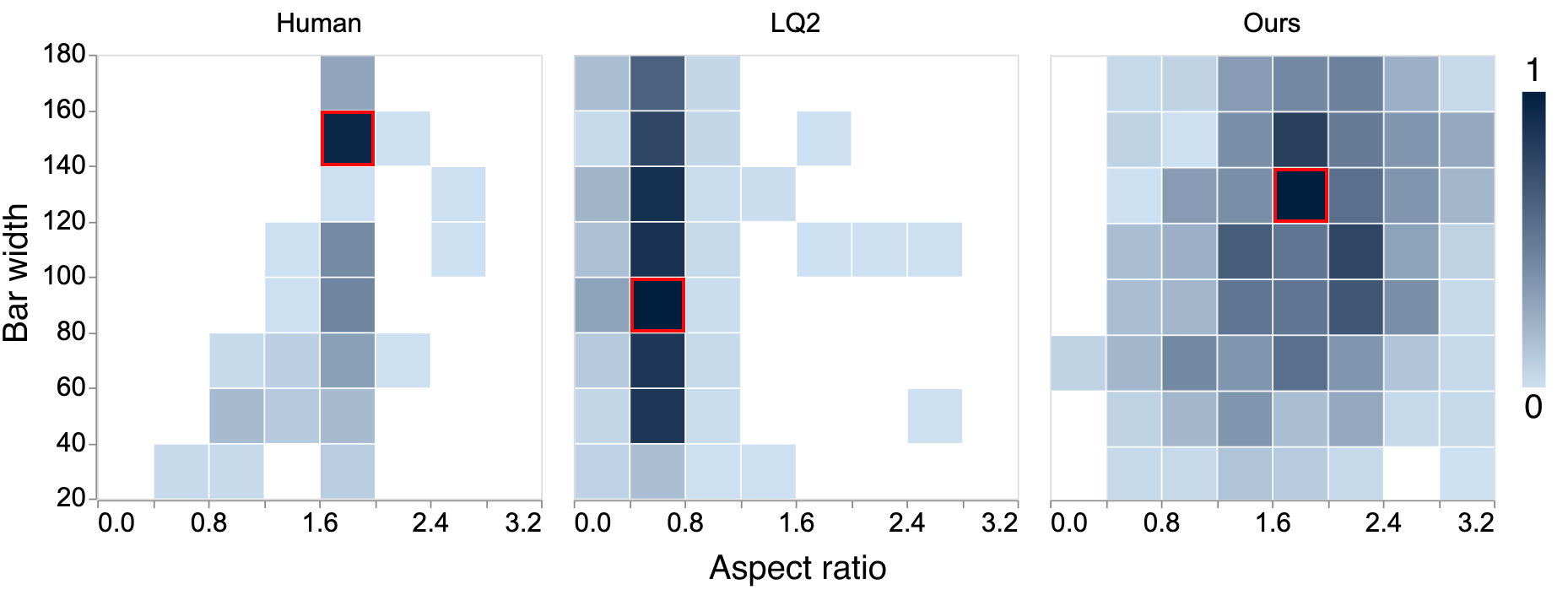}
    \caption{Heatmaps of the normalised distribution of human-selected parameters and parameters optimised using LQ2 and \methodname across 300 charts. The figure shows the distribution for the only two design parameters LQ2 can optimise: aspect ratio and bar width. The most commonly used parameter combinations are highlighted in red. \methodname picks parameters that closely align with those the human experts select. In contrast, LQ2 deviates notably from the human parameters.}
    \Description{The heatmaps display the normalised distribution of human-designed parameters and optimised parameters from our method and LQ2 across 300 charts.}
    \label{fig:parameter-analysis}
\end{figure}

We further investigated the optimised parameters in the design space and compared our \methodname with both human-designed charts\cite{masry2022chartqa} and LQ2-optimised designs~\cite{wu2021learning} on the 300 charts introduced in Sec. 4.1.
We selected aspect ratio and bar width as the parameters for analysis, as they are the most general design factors when humans create a bar chart~\cite{wu2021learning}. %
Figure~\ref{fig:parameter-analysis} illustrates the relationship between bar widths and aspect ratios across the 300 charts.
In the human-designed charts, the most preferred aspect ratio is between 1.6 and 2.0, with a bar width of between 140 and 160 pixels. Our \methodname closely aligned with these human-designed results, with the most frequent region having an aspect ratio of 1.6 and 2.0 when the bar width is between 120 and 140 pixels.
However, the baseline approach LQ2 yielded substantially different results from those of humans, with an aspect ratio of 0.4 and 0.8 when the bar width is between 80 and 100 pixels.
In addition, we find that \methodname covers the parameter space more comprehensively than human and LQ2. This shows that our design can be adapted to diverse design combinations.

\subsection{Other Chart Types}

\edit{
With minor modifications, \methodname could be adapted to other chart types. 
For instance, \citet{micallef2017towards} employed exhaustive search with a perceptual cost function to optimise scatterplots. 
In this context, exhaustive search could be replaced with Bayesian optimisation to more efficiently explore the continuous design space, such as the colour and opacity of data points. 
Here, we further demonstrate the application of \methodname to multi-column bar charts and circular charts. 
The following figures generated by GPT-5.1 were all generated by the same prompts as described in~\autoref{ssec:studydesign} in a temporary session.
}

\paragraph{\edit{Multi-column Bars}}

\begin{figure}[t]
    \centering
    \includegraphics[width=\linewidth]{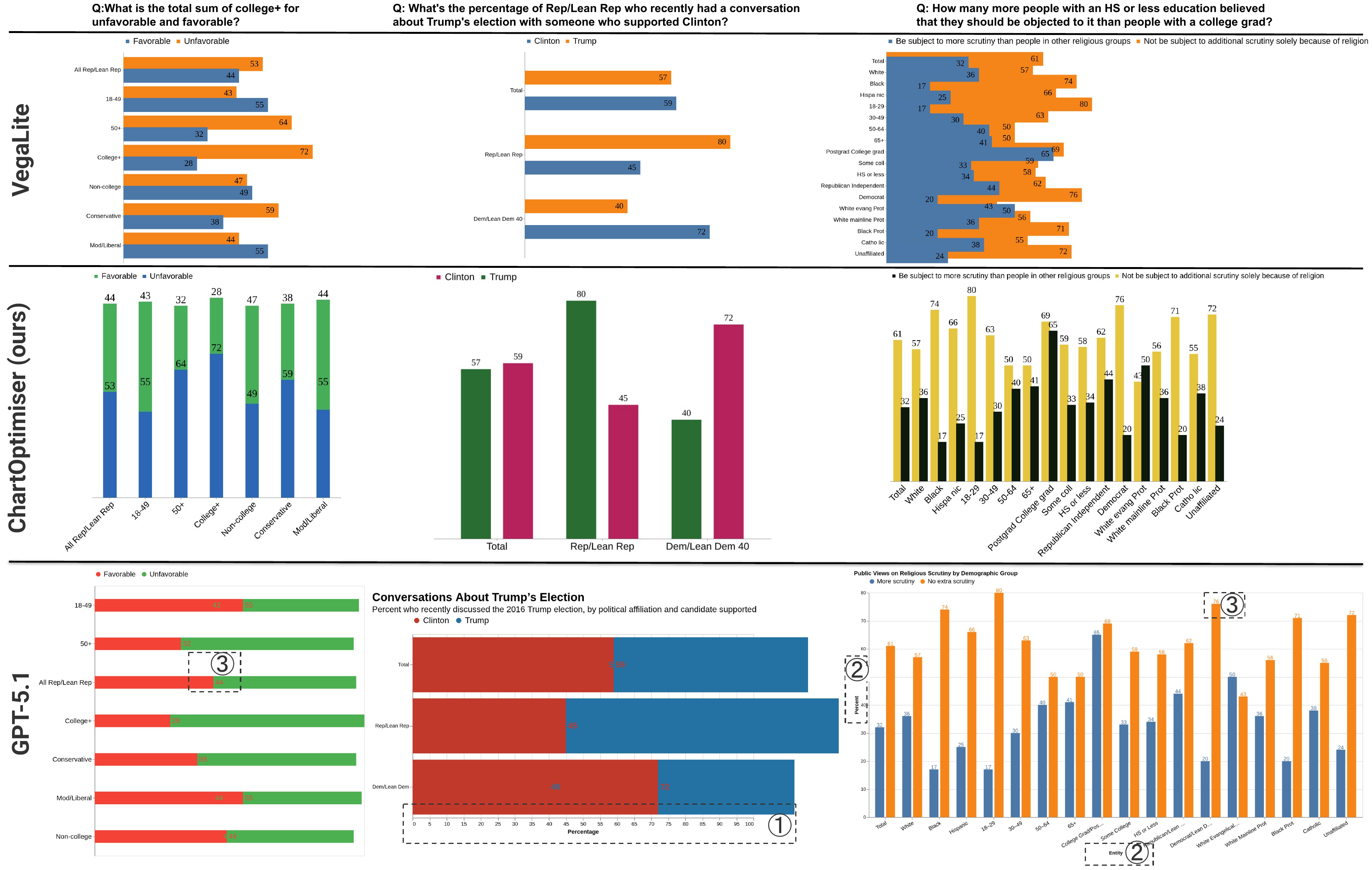}
    \caption{\edit{Comparison of automated design approaches for multi-column bar charts. Top: VegaLite defaults (Source: ChartQA). Middle: \methodname results. Bottom: GPT-5.1 outputs. Although GPT-5.1 exhibits semantic sensitivity, its outputs often suffer from design confabulation and violations of best design practices.}}
    \Description{Sample multi-column bar charts from the ChartQA dataset, and optimised bar charts from \methodname and GPT-5.1.}
    \label{fig:multicol_example}
\end{figure}

\edit{
To adapt \methodname, we expanded the design space by introducing a binary variable that indicates whether the chart is stacked. In addition, group colours were derived from the bar and highlight colour specifications to keep minimal changes to the design space. 
Weights of the objective function is set to $w_w = 3$, $w_c = 2$, $w_t = 2$, $w_s = 4$. 
Sample multi-column bar charts generated from \methodname and GPT-5.1 were illustrated in \autoref{fig:multicol_example}. 
} 

\edit{
GPT-5.1 exhibits robust capabilities for context-aware design, particularly in encoding group semantics through colour, e.g. mapping Clinton (Democrats) to blue and Trump (Republicans) to red. 
However, the model frequently generates confabulated designs~\cite{siontis2024chatgpt} and deviates from established best practices. 
As shown in \textcircled{1}, the model incorrectly employed a stacked bar chart where the cumulative percentage exceeded 100\%, whereas a non-stacked design would have been topologically appropriate. 
Furthermore, the model tends to violate the principle of minimalism by introducing visual clutter; \textcircled{2} illustrates the addition of redundant axis labels that diminish the data-ink ratio without providing auxiliary information. 
Finally, in an attempt to balance the layout \textcircled{3}, GPT-5.1 reduced font sizes to 16--18 px, thereby severely compromising readability.
On the contrary, \methodname consistently yields chart designs that prioritise structural coherence and a high data-ink ratio. Although it lacks the semantic colour intuition of LLMs, it ensures high-level readability through optimised spacing, appropriate font sizes, and maintains data integrity, leading to efficient and effective chart designs.
}

\paragraph{\edit{Circular Charts (Pie and Doughnut charts)}}

\begin{figure}[t]
    \centering
    \includegraphics[width=\linewidth]{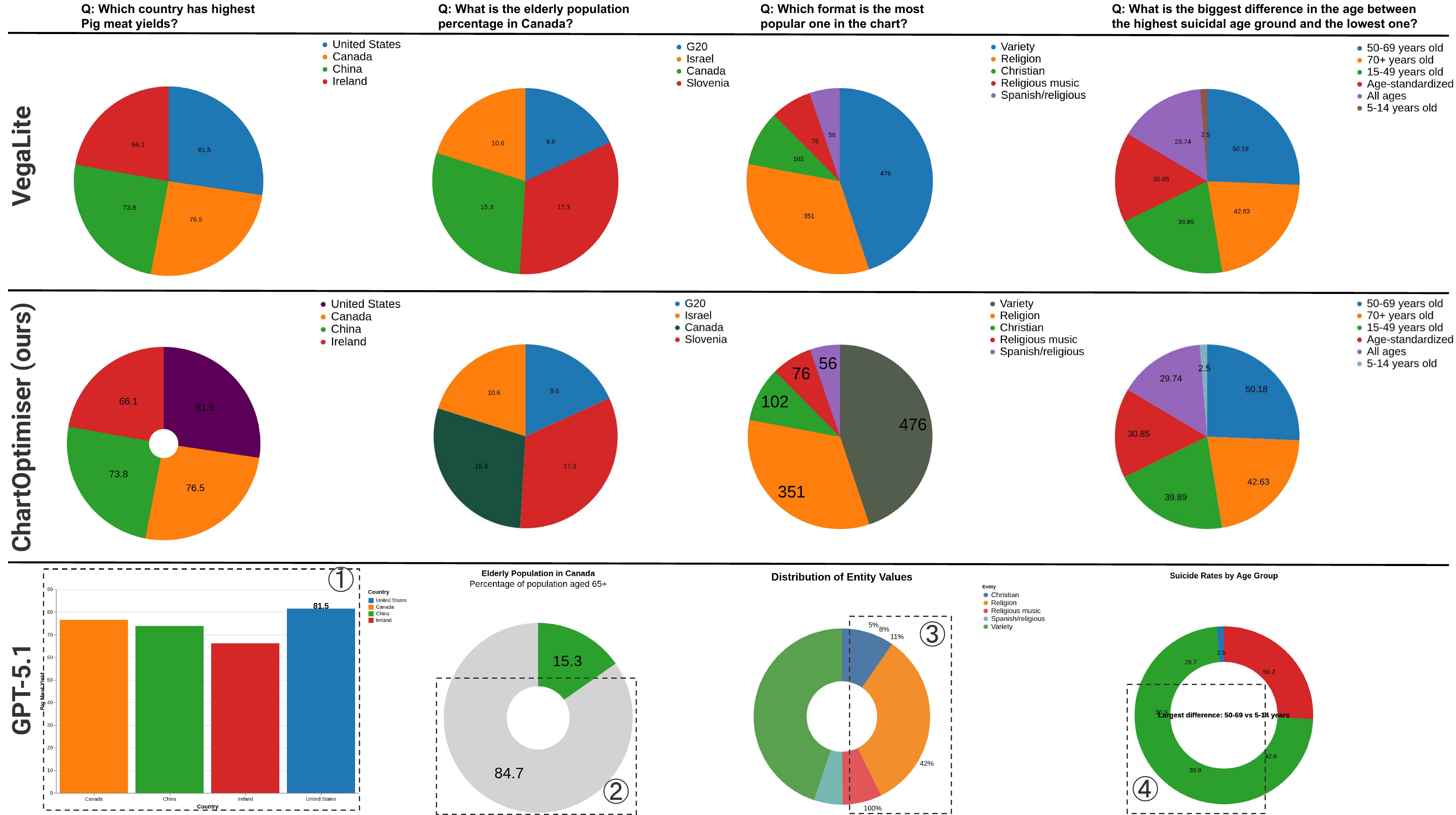}
    \caption{\edit{Comparison of automated design approaches for circular charts. Top: VegaLite defaults (Source: ChartQA). Middle: \methodname results. Bottom: GPT-5.1 outputs.}}
    \Description{Sample circular charts from the ChartQA dataset, and optimised bar charts from \methodname and GPT-5.1.}
    \label{fig:circular_example}
\end{figure}

\edit{
The pie charts were converted from the VegaLite default bar charts described in \autoref{ssec:studydesign}, with sorted data in descending order corresponding to the colour palette Category10. 
For circular charts, \methodname has font size (data), highlight data colour, inner radius, and the text position (radius) in the design space. 
Weights of the objective function is set to $w_w = 0$, $w_c = 1$, $w_t = 1$, $w_s = 1$, removing the white space ratio from the function. 
We added one more constraint to make sure the data labels always appear within the doughnut. 
Sample circular charts generated from \methodname and GPT-5.1 were illustrated in \autoref{fig:circular_example}. 
}

\edit{
GPT-5.1's output demonstrates several instances of design confabulation. First, the model \textcircled{1} arbitrarily altered the visualisation topology to bar charts while failing to annotate the full dataset. 
Second, data fidelity was compromised through \textcircled{2} the incorrect aggregation of distinct data entries. 
Moreover, the model failed to adhere to standard encoding heuristics, evidenced by \textcircled{3} erroneous data transformations (e.g., converting raw values to percentages where $100\%$ indicates a clear logic failure) and the spatial displacement of textual labels. 
Additionally, \textcircled{4} the model removed the legend while assigning an identical green colour to multiple data categories, creating ambiguity. 
In contrast, \methodname strictly maintains data fidelity. It enhances visual hierarchy by selectively tuning a single colour to increase the saliency of highlighted data points, while supporting switching types between doughnut and pie charts.
}

%% file: sections/discussion.tex
\subsection{On \methodname}
\paragraph{Advantage of Bayesian optimisation.}
\edit{We employ Bayesian Optimisation due to its superior performance in black-box optimisation scenarios typical of visualisation design. 
Unlike evolutionary algorithms and other combinatorial optimisations that rely on exhaustive sampling, BO's sample-efficient nature makes it highly suitable for chart design tasks. 
We further distinguish our approach from Reinforcement Learning, which typically treats optimisation as a step-by-step sequence; conversely, \methodname optimises the design space holistically. 
Additionally, \methodname overcomes a primary limitation of contemporary deep learning models: the reliance on curated training data. By utilising explicit, perceptually-grounded objective functions, our method provides a ``cold-start'' capability that ensures effectiveness even in the absence of large-scale training data.
}

The core innovation of \methodname is the proposed objective function, including four essential perceptual metrics in the Bayesian optimisation pipeline: white space ratio, colour preference, visual saliency in the highlight region, and text legibility. 
This combination enables the generation of optimised bar charts highly attuned to users' analytical tasks. The empirical results demonstrate that, by leveraging these perceptual metrics, \methodname generates charts that are competitive with human-designed charts regarding clarity and task-solving ease. 
This is particularly notable in complex tasks like computing derived values or comparing multiple data points.

\edit{
While this work primarily evaluates \methodname on bar and circular charts, the underlying framework is designed to be generalisable to a broad spectrum of chart types such as scatter and line plots. 
This generalisability stems from the usage of a high-level VegaLite grammar for the design space and the reliance on image-based perceptual metrics for the objective function.
Moreover, the design space has advanced to continuous choices, such as for colour and aspect ratio.
} 
In stark contrast, previous chart optimisation works limited their design space to discrete values~\cite{micallef2017towards, wu2021learning}.
Despite the expanded design space in \methodname, as shown in \autoref{fig:parameter-analysis}, the parameter selections closely align with human-generated designs.
\edit{
Despite this vastly expanded search space, our analysis (see \autoref{fig:overall-rating}) confirms that \methodname efficiently converges on parameter selections that closely align with professional human-generated designs. 
This capability positions \methodname as a robust solution for upgrading the suboptimal default settings in tools like Excel and VegaLite, addressing the pervasive default effect~\cite{wu2021learning} where users often unknowingly settle for adequate but ineffective visualisations.
}

\paragraph{Bayesian Optimisation versus Large-Language Models.}

\edit{
While Large-Language Models (LLMs) like ChatGPT have demonstrated impressive performance in generating high-quality images, the results in chart design highlight critical issues in fidelity, efficiency, and effectiveness. 
First, the most significant limitation of LLMs lies in their probabilistic nature. As observed in qualitative results (see \autoref{fig:multicol_example}), LLMs prioritise good-looking designs over data fidelity, leading to confabulations where the generated chart misrepresents the underlying data values. 
In contrast, \methodname operates within a strictly defined parameter space. By treating chart design as a constrained optimisation problem rather than a token prediction task, our approach guarantees data integrity.
Second, LLMs could offer a zero-shot and quick solution, but achieving high-quality results with LLMs often necessitates expert-level prompt engineering to correct design flaws, resulting again in a human-in-the-loop process. 
It is a chicken-and-egg problem that users require to have design knowledge to make LLMs generate satisfying chart designs. 
Third, while LLMs rely on learned correlations from training data to approximate plausible design, they lack an explicit understanding of human visual perception. This results in charts that may be aesthetically pleasing but functionally deficient for specific tasks. 
\methodname bridges this gap by directly maximising a rigorous objective function grounded in perceptual metrics (saliency, colour preference, text legibility, and visual density). This ensures that the resulting designs are not just valid code, but are empirically effective at guiding user attention to the most relevant information for their specific analytical goals.
}

\edit{
We believe that \methodname offers a valuable complement to LLMs, specifically thanks to its fidelitous, effective, and efficient performance.
First, the BO process has eliminated human-driven iteration (prompt engineering) and is more efficient compared to GPTs.
Second, we noticed that GPT modified data entries in some cases, which raises an issue regarding the validity of the results. 
}

\subsection{On the User Study Results}

The user study revealed essential insights into the performance of \methodname. 
Participants consistently rated \methodname highest in clarity and task-solving ease across all task conditions, demonstrating the value of incorporating perceptual metrics like saliency and text legibility in optimising chart designs (see \autoref{fig:overall-rating}). 
The four most frequently mentioned factors -- highlighted key data points, clear labelling, a pleasing colour scheme, well-spaced bars, and legible font sizes -- strongly aligned with our objective function, confirming \methodname’s effectiveness in enhancing visual clarity and usability.

Notably, \methodname demonstrated strong performance in complex tasks (CP and CDV), likely due to the method’s effective layout design and colour usage to emphasise task-related data points (see \autoref{fig:task-rating}).
In contrast, for simple tasks like RV and FE, \methodname did not achieve the first place. 
This finding is consistent with prior work suggesting that perceiving extremum values requires minimal cognitive effort~\cite{elzer2011automated}. 
The insignificant differences were observed in the FE tasks, indicating those simple tasks do not necessarily require an optimised design for the observation condition.

Moreover, human-designed charts significantly outperformed all automatic methods in terms of aesthetics, which indicates that while automation can greatly assist with functional aspects, human designers can still better capture the subtle visual elements that enhance the overall appeal of a chart. 
Nevertheless, \methodname ranked second in aesthetics, promising to balance functionality and visual appeal.

\subsection{On Applications}

\paragraph{Accessibility}
Data Visualisation Literacy is the ability to interpret visual patterns in the visual domain as properties in the data domain~\cite{boy2014principled}.
Users with low Data Visualisation Literacy may need help with complex data interpretation tasks, such as comparison or computing derived values. \methodname has strong competence in optimising charts for those complex tasks, making the charts more engaging.
For users with low visual literacy, complex visualisations can be overwhelming. To mitigate this, \methodname could prioritise lower information density by increasing the white space ratio, thus avoiding clutter and presenting only essential information. This would help users focus on the most critical aspects of the data.
The colour design space of \methodname could be simplified to ensure accessibility. Prioritising contrasting and accessible colour schemes would facilitate quicker comprehension for users with low visual literacy~\cite{borner2019data}. Moreover, ensuring that charts adhere to colour guidelines for colour blindness, like avoiding red-green contrasts, makes visualisations more inclusive and comprehensible to a broader audience.

\paragraph{Content Localisation}

\begin{figure}[t]
    \centering
    \includegraphics[width=0.78\linewidth]{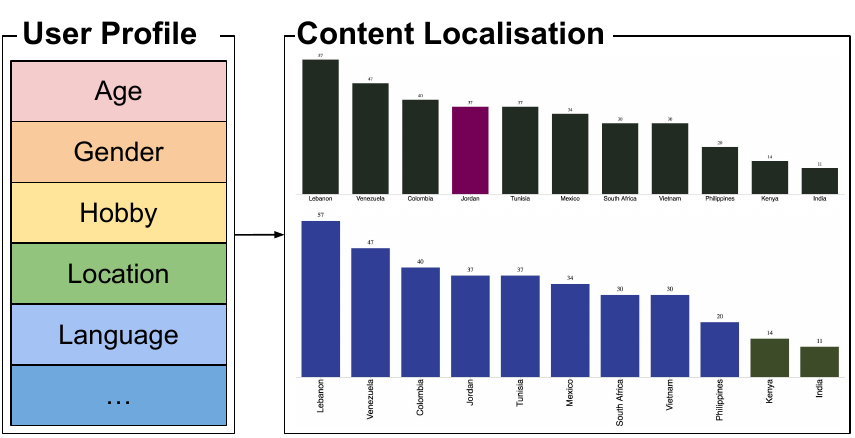}
    \caption{Sample content-localised charts generated by our method based on a user profile. The chart at the top was optimised for a user in Jordan, while the lower one was for a user interested in Kenya and India.}
    \Description{Sample content-localised charts generated by our method based on a user profile.}
    \label{fig:application}
\end{figure}

Journalists and media outlets frequently use visualisations to convey data to the public. 
Using \methodname can enhance these visualisations by adaptively emphasising different data points for different audiences, a process known as content localisation~\cite{kee2015review}. 
\autoref{fig:application} visualises the percentage of people accessing a personal computer in different countries. For instance, the top figure is relevant for users located in Jordan, as Jordan is highlighted in the chart. The bottom figure fits the user with interest (e.g. travel plans) in Kenya and India. 
\methodname ensures that the charts are visually appealing and optimises them for relevance to specific audiences based on their demographic information (user profile), making it easier for readers to extract key insights and increase engagement with localised stories.
Nevertheless, more engineering work is needed given localisation differences, such as reversing the order of axis labels given a language reading from right to left, and choosing culturally plausible colour schemes.

\paragraph{Thumbnailing and Grey-scale Charts}
\edit{Beyond standard displays, \methodname can be adapted to highly constrained environments, such as thumbnails and print media. In UI applications like dashboards~\cite{deng2022dashbot, jiao2010visual}, size reduction necessitates prioritising immediate recognisability over fine-grained detail. 
\methodname could automatically emphasise critical data features while eliding illegible elements like legends or minor axis labels. 
Similarly, adapting visualisations for greyscale print media (e.g. newspapers) removes colour as an encoding channel. In such scenarios, our system could be reconfigured to compensate for the loss of colour by optimising alternative parameters -- such as aspect ratio, bar width, or hatching -- to maintain structural coherence and communicative intent under technical limitations.
}

\subsection{Limitations and Future Work}
\edit{
A significant limitation of the current \methodname implementation is the requirement for manual specification of the design space and optimisation goal tailored to specific chart types. This manual intervention restricts its practical ability. 
To enhance \methodname's versatility, future work should focus on automatically identifying the chart type. 
This would allow the method to adapt its internal optimisation parameters autonomously, streamlining the generation process. 
Furthermore, the current method's chart coverage is constrained. 
It is challenging for \methodname to accommodate the full spectrum of chart types used in practice, such as area charts and map visualisations. 
Leveraging the contextual memory provided by LLMs represents a viable path toward greater sophistication. 
This integration would enable \methodname to generate dynamic and interactive charts that are optimised to flexibly support a combination of concurrent user demands.
}

\edit{
While the evaluation of \methodname employed ChatGPT to represent the state-of-the-art in generative design, its application was restricted to prompt-level JSON editing rather than a fully integrated optimisation process in a deep-learning model. 
This leaves a significant avenue for future exploration involving more systematic applications of Large Language Models. 
Future work could investigate hybrid architectures that combine the semantic reasoning of multimodal LLMs with the rigorous constraints of our optimisation engine, potentially unlocking new capabilities for end-to-end, model-driven visualisation synthesis.
}

Another avenue for future work is data storytelling in data visualisation~\cite{shi2020calliope, shao2024data}, integrating narrative-driven elements into the automatic chart design process presents a promising direction~\cite{deng2022dashbot}. 
Extending \methodname to incorporate storytelling could support a wider range of applications, particularly in dashboard design and data communication, where contextualised narratives are essential. 
Furthermore, \methodname holds the potential to be integrated into a ``ChartGPT'' system~\cite{tian2024chartgpt}, where optimised charts are dynamically generated as responses to user queries, offering tailored data insights on demand. 
This direction would enhance interactive data exploration and make data insights more accessible through conversational interfaces.

%% file: sections/conclusion.tex
In this work, we introduced \methodname\,-- the first method for task-driven optimisation of information visualisations.
At its core is a Bayesian approach that optimises chart designs with respect to a novel objective function integrating four perceptual metrics -- visual saliency, text legibility, colour preference, and white space ratio.
Through empirical evaluation of common analytical tasks, we demonstrated that our method can effectively optimise bar chart designs to enhance chart clarity, aesthetics, and ease of task-solving.
 \methodname demonstrated significant performance improvements over human-designed and baseline charts, especially in more complex tasks such as comparing or computing derived values.